\documentclass[12pt,preprint]{aastex}

\begin{document} 

\title{An Extinction Threshold for Protostellar Cores in Ophiuchus}

\author{Doug Johnstone$^{1,2}$, James Di Francesco$^{1}$, and Helen Kirk$^{2}$}

\affil{$^1$National Research Council of Canada, Herzberg Institute of 
Astrophysics, 5071 West Saanich Road, Victoria, BC, V9E 2E7, Canada; 
doug.johnstone@nrc-crnc.gc.ca, james.difrancesco@nrc-cnrc.gc.ca}

\affil{$^2$Department of Physics \& Astronomy, University of Victoria, 
Victoria, BC, V8P 1A1, Canada; hkirk@uvastro.phys.uvic.ca}

\begin{abstract} 
We have observed continuum emission at $\lambda$ = 850\,$\mu$m over $\sim$4 
square degrees of the Ophiuchus star-forming cloud using SCUBA on the JCMT,
producing a submillimetre continuum map twenty times larger than previous
Ophiuchus surveys. Our sensitivity is 40\,mJy\,beam$^{-1}$, 
a factor of $\sim$2 less sensitive than earlier maps.
Using an automated identification algorithm, we detect 
100 candidate objects. Only two new objects are detected outside
the boundary of previous maps, despite the much wider area surveyed.
We compare the submillimetre continuum map with a map of visual 
extinction across the Ophiuchus cloud derived using a combination of 2MASS 
and $R$-band data.  The total mass in submillimetre objects
is $\approx$ 50 M$_\odot$ compared with $\approx$ 2000 M$_\odot$ in 
observed cloud mass estimated from the extinction. The submillimetre
objects represent only 2.5\% of the cloud mass.
A clear association is seen between 
the locations of detected submillimetre objects and high visual extinction, 
with no objects detected at $A_{V}$ $<$ 7 magnitudes. 
Using the extinction map, we estimate 
pressures within the cloud from $P/k \approx 2\times 10^5\, {\rm cm}^{-3}\, 
{\rm K}$ in the less-extincted regions to $P/k \approx 2\times 10^6\,{\rm 
cm}^{-3}\, {\rm K}$ at the cloud centre.  Given our sensitivities, cold ($T_d 
\approx 15\,$K) clumps supported by thermal pressure, had they existed,
should have been detected throughout the majority of the map.  Such objects
may not be present at low $A_{V}$ because they may form only where 
$A_{V}$ $>$ 15, by some mechanism (e.g., loss of non-thermal support).  
\end{abstract}


\section{Introduction}

Stars must form out of gravitationally bound substructures within a molecular 
cloud but how the substructures themselves form is strongly debated.  Possible 
scenarios for this process range from the gradual release of magnetic support 
\citep{MS56,S83,N84,SAL87,MC99,BC04}, leading to a slow, regulated evolution of 
structure in the cloud, to violent dissipation through interacting waves 
(Scalo 1985; Mac\ Low \& Klessen 2004 and references therein).

Rotational transitions of molecules have been important for tracing both
substructures and motions within molecular clouds.  Transitions that probe 
low densities in clouds (e.g., CO (1-0)) reveal widespread emission.  Lines
observed over low-density regions are themselves very wide, indicating the 
presence of non-thermal motions on large scales that are crucial to overall 
cloud support.  Transitions that probe high densities (e.g., NH$_{3}$
(1,1)), however, reveal more localized pockets within clouds, i.e., dense 
cores.  Lines observed in dense cores are narrower than those observed in
lower-density regions, indicating that non-thermal motions are reduced on 
smaller scales, with a concomitant reduction of support \citep{GB98, Gea98}.  
Dense cores are typically associated with embedded protostellar objects, 
suggesting that they are the gravitationally bound substructures from which 
stars form \citep{TMF00, DAM04}.

Although observations of transitions that probe high densities have been
a reliable means to investigate the small scale substructures of molecular 
clouds related to star formation, observations of submillimetre continuum 
emission from such regions are an attractive alternative given the high 
sensitivities such observations have to mass.  For example, large-format
bolometer arrays on large submillimetre or millimetre telescopes have 
been used in recent years to map several pc$^{2}$ of nearby star-forming 
molecular clouds, revealing many substructures in high-density regions at 
high sensitivities and resolution \citep{MAN98,JB99,JWM00a,JFM01}.  In these
regions, the detected continuum emission primarily originates from cold dust grains
and is almost always optically thin, so continuum maps reveal 
temperature-weighted column densities.  These data have revealed numerous 
stellar-mass 
substructures within the molecular cloud cores whose mass distributions 
have similarities to the stellar initial mass function.  

Most of the regions targeted for widefield submillimetre continuum mapping
have been those known previously to contain dense cores and 
active star formation.  Much of the surrounding regions of lower-density 
cloud have been ignored in these maps,
in part to maximize sensitivities.  Significant information about the 
development of substructures may exist in these regions, however, but has
been overlooked as a result of this bias. For example, investigation of Orion
B North \citep{JFM01} found all the substructure to be confined to
the densest molecular core regions but did not have sufficient
cloud coverage to provide explicit environmental constraints.
To provide an unbiased view of
substructures within nearby star-forming molecular clouds, we present here
submillimetre continuum and infrared extinction maps of $\sim$4 square 
degrees of the Ophiuchus star forming cloud.  These observations are part 
of a larger project, the CO-ordinated Molecular Probe Line, Extinction, and
Thermal Emission (COMPLETE) project \citep{G04}, 
to obtain and compare high quality 
molecular line, submillimetre continuum, and infrared extinction data over 
the extents of the nearby Ophiuchus, Perseus, and Serpens star-forming 
molecular clouds.  (These clouds, as well as the Lupus and Chamaeleon clouds
and many isolated cores, being observed extensively as part of the near-
to mid-infrared ``From Cores to Disks" (c2d) Legacy survey of the Spitzer 
Space Telescope \citep{Eea03}.)  The data presented here represent about
half of the entire extent of Ophiuchus that will be mapped in the
submillimetre for COMPLETE but
already they comprise a map that is an order of magnitude larger than those 
of Ophiuchus by \citet{MAN98} or \citet{JWM00a}
(albeit at lower sensitivity) and they sample an order of magnitude in 
column density of the cloud.

In this Letter, we demonstrate that all substructures detected in the 
Ophiuchus cloud are confined to a small fraction of the submillimetre
continuum map.  Despite the tenfold increase in the area of Ophiuchus 
probed here, only two objects located off the edges of previous 
maps are detected.  No obvious substructures are found 
below an $A_{V} = 7$.  Moreover, almost all bright submillimetre objects 
are found in the most-opaque, densest part of the cloud where $A_{V}$
$\geq$ 15 mag although simple models suggest they would have been detected 
at our sensitivity throughout most of the map if they existed.  Instead, 
a column density or extinction threshold for substructure formation, and 
thus star formation, may exist in Ophiuchus and possibly in other 
molecular clouds.

\section{Observations and Results}

The data were 
obtained using the Submillimetre 
Common User Bolometer Array (SCUBA; see \citet{Hea99}) on the 
James Clerk Maxwell Telescope (JCMT) on Mauna Kea, Hawaii\footnote{The 
JCMT is operated by the Joint Astronomy Centre in Hilo, Hawaii on behalf 
of the parent organizations Particle Physics and Astronomy Research Council 
in the United Kingdom, the National Research Council of Canada and The 
Netherlands Organization for Scientific Research.}.  The area chosen to 
map consisted of $\sim$4 square degrees in western Ophiuchus where 
$A_{V}$ $>$ 3 and includes the well-known cores Oph A, B1, B2, C, 
and D at its southeastern corner.  This area was divided up into 38 square fields 
each $\sim$400 square arcminutes in size.  Each field was observed over 
the course of $\sim$1 hour using the ``fast scan" observing mode of SCUBA 
where the sky is sampled at a Nyquist rate at 850 $\mu$m.  Each field was 
scanned six times but at a different chopping amplitude and direction each 
time. The chop maps were converted into an image by applying the matrix
inversion data reduction technique \citep{JWM00b}. 
Note that by using a chopping secondary to map the region, the data 
are insensitive to emission on scales larger than a few times the chop throw, 
i.e., $>$120\arcsec.  The fields were observed separately through queue mode 
scheduling throughout Semester 03A.  Since each field was observed at 
different times and thus under different atmospheric opacities, the 
sensitivity in the final map varies from field to field but only by a 
factor of $\sim$1.5.  The mean and rms of the 1 $\sigma$ rms sensitivity 
for all fields is 40 mJy beam$^{-1}$ and 20 mJy beam$^{-1}$ respectively.
Since the JCMT beam is $\sim$14\arcsec\ FWHM at 850 $\mu$m, the final 
map consists of $\sim$1.9 $\times$ 10$^{5}$ resolution elements, the
largest number obtained so far in a single SCUBA map.

Extinctions derived from $R$-band star count data by \citet{Ca99} were used 
for the northern half of the map while those derived from 
Two Micron All Sky Survey (2MASS; \citet{Sea97}) stellar reddening 
data by \citet{Aea04}
as part of COMPLETE were used for the southern half since public release 
data were available for those fields.  Since 2MASS is an infrared survey, 
its data are better for probing the highly-extincted regions at which
the $R$-band extinctions saturate, $A_{V}$ $\approx$ 8.  
Extinctions derived from 
$R$-band or 2MASS data over regions in common 
show a slight difference of less than 1 mag and a tight linear correlation 
until $A_{V}$ $\approx$ 8.  
The resolution of the extinction map depends on the densities of stars used 
to derive extinctions, 
but it should be only as low as $\sim$3\arcmin\ at the highest extinctions
(see Lombardi \& Alves 2001).  The mean and maximum 
$A_{V}$ over the surveyed area are 4.2 and 36.2 mag respectively.

\section{Discussion}

The $\lambda$ = 850 $\mu$m continuum map of Ophiuchus can be used to locate 
the incidences of substructure (i.e., objects of size $< $120\arcsec) in the
cloud.  To locate objects within the map in an unbiased manner, the Clumpfind 
algorithm \citep{WGB94} was used to identify 100 sources of flux in the 
map of comparable size to the beam or larger to a 
5 $\sigma$ threshold in peak brightness.  All but 2 of these objects are
associated with previously identified substructure in the smaller map 
of \citet{JWM00a} surrounding the known Oph cores (i.e.,
A--G; see also \citet{MAN98})  Table 1 lists the names, positions, and fluxes 
of two objects not previously identified since they lie beyond 
the edges of earlier continuum maps. They reside in two new cores in 
Ophiuchus, which we name Oph H and Oph I following \citet{MAN98}.  
No infrared or X-ray sources toward either core are reported
within the SIMBAD database.

Detections of substructure in Ophiuchus allow us to measure their mass 
and determine their location while the extinction map can be used to 
estimate the total cloud mass over the same area.  The integrated
flux from all substructure in the cloud is 250 Jy. Assuming a dust 
temperature $T_{d}$ = 15 K, a dust opacity at 850 $\mu$m 
$\kappa_{\nu}$ = 0.02 cm$^{2}$ g$^{-1}$, and a distance of 160 pc
to Ophiuchus then 1 Jy = 0.20 M$_{\odot}$ \citep{JWM00a}. Therefore, 
the amount of mass in substructures detected is $\sim$ 50 M$_{\odot}$.  
From the extinction data and 
assuming $A_{V}$/N$_{H}$ = 2 $\times$ 10$^{21}$, the total amount of mass 
residing within the map area is $\sim$ 2020 M$_{\odot}$.  The detected 
substructures, therefore, only account for $\sim$2.5\% of the total amount of 
mass residing in the map area.  Table 2 shows the distribution of these masses 
with $A_{V}$, and reveals that significant mass in substructure is found only 
at $A_{V}$ $>$ 15 mag. 

\begin{figure}
\plotone{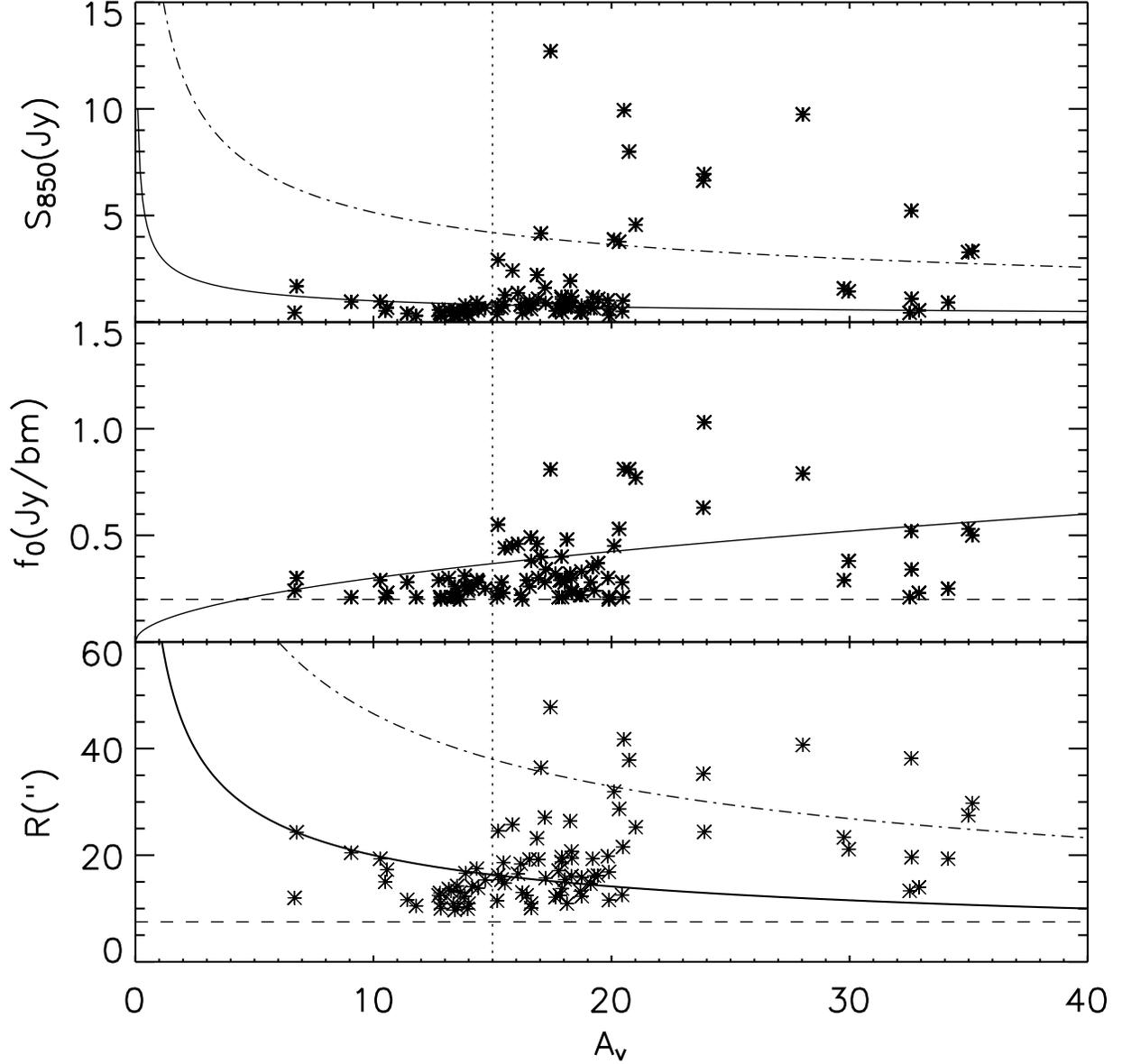}
\vskip -0.75in
\caption{ 
Distribution of the observed properties of measured submillimetre clumps
with cloud extinction. Panels from top to bottom show for each clump the
integrated flux, peak flux, and size. Horizontal dashed lines denote
sensitivity and resolution limits while vertical dotted lines at $A_{V}$ = 
15 mag mark the boundary of observed bright objects. The solid
lines show the expected evolution with $A_{V}$ of a BE sphere with
a fixed value of $\lambda$  but only marginally observed at
$A_{V}$ = 10 mag (see text). The dash-dotted lines show the upper limit
for size and mass of stable BE spheres ($T = 15\,$K) as a function of
$A_{V}$.
}
\label{fig:src}
\end{figure}

Cumulative mass calculations may be easily biased by a few 
massive objects, but this is not the case for our surveyed area.  Figure 
\ref{fig:src} plots the peak flux, total flux, and radius of each
identified object versus the $A_{V}$ at their respective positions.
No bright objects are found at $A_{V}$ $<$ 15 despite the large cloud area 
and mass of that region.
A similar result was obtained by \citet{JFM01} for submillimetre
objects in Orion B North.
At high extinction the total mass in clumps
is determined not by a single object but by an ensemble, although the Oph A
core accounts for one quarter of the mass in compact objects. Comparison of 
the mass estimation in this paper with the results of \citet{JWM00a} shows 
that a significant amount of mass (almost 50\% of the submillimetre flux in 
compact substructure) in the earlier paper was removed by the flattening and 
filtering techniques, stressing the difficulty in obtaining proper flux 
baselines for chopped submillimetre maps.

What is the physical significance of having most of the detectable 
substructure only at $A_{V} > $ 15 mag? A threshold in $A_{V}$ may need to be 
exceeded for such objects to form. Alternatively, our observations may not 
have been sensitive enough to detect substructure in regions of low $A_{V}$, 
e.g. if such objects are larger or less dense than their more-embedded 
brethren. Here, we use a simple model to argue that such objects in locations 
of low $A_{V}$ in Ophiuchus should have been detected by our observations if 
they indeed existed.

Within a molecular cloud, pressure increases with $A_{V}$.  Following 
\citet{M89}, the pressure at depth $r$ is 
$P(r) = \pi\, G\, \bar{\Sigma}\, \Sigma(r)$,
where $\bar{\Sigma}$ is the mean column density through the cloud and
$\Sigma(r)$ is the column density measured inward from the cloud surface 
to depth $r$.  Near the centre of the cloud, $\Sigma(r) \approx \Sigma(s)/2$,
where $\Sigma(s)$ is the column density through the cloud at impact
parameter $s$.  Thus, pressure in the cloud at depth $r$ can be 
approximated as
$P(r)/k = 1.7 \times 10^4\,\bar{A_{V}}\,A_{V}(s)\ {\rm cm}^{-3}\,{\rm K}$,
where $\bar{A_{V}}$ is the mean extinction through the cloud, $A_{V}(s)$ is 
the extinction through the cloud at impact parameter $s$, and adopting
$A_{V} = (\Sigma / \Sigma_0)\,$mag where $\Sigma_0 = 4.68 \times 10^{-3}\, {\rm g}\,{\rm cm}^{-2}$.
Taking the mean extinction through the cloud as $\bar{A_{V}} \approx 4$ and 
the range of measured $A_{V} = 3 - 30\,$ mag, $P/k \approx 0.2-2 \times 
10^6\,{\rm cm}^{-3}\,{\rm K}$.  Although admittedly approximate, the 
value at the highest extinctions (e.g., the Oph cores) is similar to the 
$3 \times 10^6\,{\rm cm}^{-3}\,{\rm K}$ found by \citet{JWM00a} from fitting
stable Bonnor-Ebert (BE) sphere models \citep{B56,E55}
to submillimetre objects in the Ophiuchus
cores to estimate their masses, internal temperatures, and bounding pressures.

Pressure support for a molecular cloud must come from non-thermal sources like
magnetic fields or turbulence.  Most of the previously-detected submillimetre 
objects in Ophiuchus, however, were fit reasonably well by models of BE 
spheres supported internally entirely by thermal pressure 
\citep{JWM00a}. A similar analysis holds true for Orion B North \citep{JFM01}.
Note that these fits should not be construed as evidence that the objects
are in stable equilibrium, since dynamic entities produced in turbulent 
clouds can appear like BE spheres \citep{BKV03}. Similarly, the
condensed central regions of clumps evolving via ambipolar diffusion approach
a similar, albeit flattened, profile \citep{B97,BC04}.  
The BE sphere model, however,
provides a lower limit for the critical mass and column density at which 
an object will become unstable to gravitational collapse since only thermal 
pressure is available to act as support.  Aside from scaling relations, BE
spheres are a one-dimensional family of objects defined only by the ratio
of central density to surface density, $\lambda =\rho(0)/\rho(R)$, 
with larger $\lambda$ implying a greater importance of self-gravity. 
BE spheres are stable against gravitational collapse if $\lambda < 14$.
The mass and density scaling relations of a BE sphere with fixed
$\lambda$ are $M_{BE} \propto P^{-1/2}$ and 
$\Sigma_{BE} \propto P^{1/2}$.

Can stable objects exist in Ophiuchus where $A_{V}$ $<$ 15 mag and remain
undetected in our data?  The total submillimetre continuum flux $S_{850}$
associated with a BE sphere is a measure of its mass, weighted by 
temperature.  Thus, for BE spheres with a fixed value of $\lambda$, the
submillimetre flux as a function of the bounding pressure (assuming that both 
$\kappa_\nu$ and $T$ remain constant) is $S_{850} \propto P(r)^{-1/2}$ or 
$S_{850} \propto A_{V}(s)^{-1/2}$. Similarly, the peak flux $f_0$ measures the
column density and should scale as $f_0 \propto A_{V}(s)^{1/2}$ and
the radius $R \propto A_{V}(s)^{-1/2}$.  The total flux, peak flux, and
radius {\it increase} with increasing $\lambda$ 
(until a gravitationally unstable BE sphere is produced).

In Figure \ref{fig:src} the above BE sphere relations are overlaid (solid
line) on the data to show that non-detections at $A_{V} = 10$ are 
incompatible with detections at higher $A_{V}$. That is, the lack of 
significant observed substructure at moderate $A_{V}$ 
implies that no objects 
in these regions are dominated by self-gravity to the extent of those at 
higher extinction.  Since the low $A_{V}$ regions contain most of the cloud 
mass, the lack of such detected objects is not due to a lack of material. 
Also shown in Figure  \ref{fig:src} is the upper boundary for stable BE 
spheres (dashed-dotted line), assuming that the gas and dust temperature is
$T = 15\,$K.  Unstable objects, those which are capable of collapsing to stars,
are only found where $A_{V} > 17$.

How do these results fit with present theoretical models for star formation?
If the dominant mechanism for molecular cloud support is magnetic pressure
mediated by collisions between ions (which couple to the magnetic field)
and neutrals then the timescale for ambipolar diffusion would be long
in regions of high ionization, limiting the formation
of significant substructure. \citet{Cr99} has shown that the 
typical magnetic field strength in a molecular cloud is strong enough to
supply most, if not all, of the required support against collapse.
\citet{M89} demonstrates that in the
lower column density regions of molecular clouds, where the extinction to the
cloud surface is $A_{V} < 4$, the interstellar radiation field is capable of 
maintaining such a high ion fraction. Following this reasoning, the total
extinction through regions of the cloud where substructure exists
should be $A_{V} > 8$. This prediction is in excellent agreement with our 
observations. The somewhat higher threshold $A_{V}$ that we find is plausibly 
a result of the non-uniform structure of the molecular cloud.
In the regions of high extinction where only cosmic
rays are capable of maintaining an ionization fraction, ambipolar diffusion 
enables the growth of prestellar cores. Recent theoretical calculations show 
that MHD turbulence may shorten the timescale for ambipolar diffusion within
the core by a factor of 3-10 \citep{FA02,Z02}, bringing the ratio of
observed cores with and without stars into agreement with their expected
lifetimes. \citet{JWM00a} found that approximately half their submillimetre
objects in Ophiuchus were associated with infrared emission\footnote{The
high sensitivity of the Spitzer c2d survey will likely change this fraction.}.

An alternative model for the support of molecular clouds utilizes turbulent
motions to supply `non-thermal' pressure support \citep{MK04}.  
Such motions are observed within molecular clouds via the widths of 
molecular transitions like CO $1-0$.  Indeed, simple virial analysis models 
indicate that the observed linewidths are large enough to provide support of 
the cloud against collapse, assuming that the `non-thermal' pressure is 
isotropic and uniform throughout the cloud \citep{L81}.
For Ophiuchus typical linewidths are $\Delta V > 1.5\,$
km\,s$^{-1}$ \citep{L89}. Within turbulent clouds,
the formation of substructure is due to 
intersecting waves, producing local density maxima, most of which are
transient. \citet{BKV03} showed that such features can look similar to BE
spheres despite being non-equilibrium objects.  It remains to be determined,
however, how the formation of substructure depends on local
conditions, such as the mean pressure. Naively one might expect that
these objects should follow a similar scaling relation to BE spheres.
Regardless of the exact nature of the scaling, it is difficult to understand 
how turbulent models can produce the sharp decline in observed objects 
found in this investigation.

Molecular clouds are likely influenced by both magnetic fields
and turbulent motions. The SCUBA observations presented in this paper suggest 
an extinction threshold, which is compatible with the predictions of 
\citet{M89}, preventing the formation of
relevant substructure over the bulk of the cloud. Thus, observations of
submillimtere structure in molecular clouds yield similar conclusions
to surveys of the latter-stage embedded stars \citep{LL03}. As with 
the embedded stars, most if not all of the submillimetre objects reside in 
highly extincted cores, account for about twenty percent of the total
core mass, and have an IMF-like initial mass function \citep{MAN98,JWM00a}.
Thus, clustered star formation appears limited to the densest regions of 
molecular clouds where turbulent motions and enhanced ambipolar diffusion 
are able to seed an ensemble of prestellar cores. 

\vskip0.1in
\noindent 
The research of D.\ J.\  has been supported by an NSERC Discovery Grant.  
H.\ K.\ is supported by an NSERC graduate scholarship. This research has 
made use of the SIMBAD database, operated at CDS, Strasbourg, France.

\clearpage 

\begin{table*}
\caption{Newly-Identified Ophiuchus Objects}
\label{tbl-1}

\begin{center}
\begin{tabular}{ccccccc}
\tableline
      &    R.A.    &    decl.  &   Peak Flux      & Total Flux &    Radius& Cloud $A_{V}$   \cr
Name  &  (J2000)   &  (J2000)  & (Jy beam$^{-1}$) &    (Jy)    & (\arcsec)&  \cr
\tableline
\tableline
H-MM1 & 16 27 58.3 & -24 33 42 &     0.40         &    4.2    &    36    & 18\cr
I-MM1 & 16 28 57.7 & -24 20 48 &     0.30         &    2.7    &    32    &  7\cr
\tableline
\end{tabular}
\end{center}
\end{table*}

\begin{table*}
\caption{Percentages of Totals in Ranges of Extinction}
\label{tbl-2}

\begin{center}
\begin{tabular}{ccccccc}
\tableline
$A_{V}$  & Cloud Area &
\multicolumn{2}{c}{Cloud Mass} &
\multicolumn{2}{c}{Clump Mass} &
Mass Ratio\cr
Range    & (\%)         & (M$_\odot$) & (\%) & (M$_\odot$) & (\%)  & (\%)           \cr
\tableline
\tableline
0--36    & 100        & 2020& 100    &	49.4& 100           &  2.5\cr
\cr
0--7    &   88        & 1380& 68     &     0&  0             &  0  \cr
7--15   &    9	      &  400& 20     &   3.1&  6             &  0.8\cr
15--36  &    3        &  240& 12     &  46.3&  94	     &  19\cr
\tableline
\end{tabular}
\end{center}
\end{table*}

\clearpage


\end{document}